\documentclass[10pt,conference]{IEEEtran}

\usepackage[utf8]{inputenc}
\usepackage{booktabs}
\usepackage{tikz}
\usepackage{cite}
\usepackage{amsmath,amssymb,amsfonts}
\usepackage{algorithmic}
\usepackage{graphicx}
\usepackage{textcomp}
\usepackage{xcolor}
\usetikzlibrary{shapes.geometric, arrows, positioning, calc}

\begin{document}

\title{Knowledge Lever Risk Management for Software Engineering: A Stochastic Framework for Mitigating Knowledge Loss}

\author{\IEEEauthorblockN{Mark Chua}
\IEEEauthorblockA{\textit{Department of Systems and Computer Engineering} \\
\textit{Carleton University}\\
Ottawa, ON, Canada \\
MarkChua@cunet.carleton.ca}
\and
\IEEEauthorblockN{Samuel Ajila}
\IEEEauthorblockA{\textit{Department of Systems and Computer Engineering} \\
\textit{Carleton University}\\
Ottawa, ON, Canada \\
SamuelAjila@cunet.carleton.ca}}

\maketitle

\begin{abstract}
Software engineering (SE) organizations operate in a knowledge-intensive domain where critical assets---architectural expertise, design rationale, and system intuition---are overwhelmingly tacit and volatile. The departure of key contributors or the decay of undocumented decisions can severely impair project velocity and software quality. While conventional SE risk management optimized for schedule and budget is common, the intangible knowledge risks that determine project success remain under-represented.

The goal of this research work is to propose and evaluate the Knowledge Lever Risk Management (KLRM) Framework, designed specifically for the software development lifecycle. The primary objectives are to: (1) recast intangible knowledge assets as active mechanisms for risk mitigation (Knowledge Levers); (2) integrate these levers into a structured four-phase architecture (Audit, Alignment, Activation, Assurance); and (3) provide a formal stochastic model to quantify the impact of lever activation on project knowledge capital. We detail the application of these levers through software-specific practices such as pair programming, architectural decision records (ADRs), and LLM-assisted development. Stochastic Monte Carlo simulations demonstrate that full lever activation increases expected knowledge capital by 63.8\% and virtually eliminates knowledge crisis probability. Our research shows that knowledge lever activation improves alignment across the project management iron triangle (scope, time, cost) by reducing rework and rediscovery costs.
\end{abstract}

\begin{IEEEkeywords}
Risk Management, Intellectual Capital, Software Engineering, LLM, Stochastic Modeling
\end{IEEEkeywords}

\section{Introduction}
The global software industry has transitioned from craft-based activity into a large-scale, distributed engineering discipline. Modern research and development (R\&D) in software is characterized by geographically dispersed Agile teams and complex microservice architectures where rapid turnover is a constant reality. In this environment, sustainable competitive advantage depends on the intellectual capital embedded within an organization's people, technical processes, and ecosystem \cite{bontis1998intellectual}.

This intellectual capital is highly vulnerable. When a lead architect departs without externalizing their understanding of system trade-offs, the project suffers ``architectural knowledge vaporization''---the loss of critical design rationale that source code alone cannot convey \cite{jansen2005software}. Standard software project risk management, such as Boehm's spiral model, provides robust mechanisms for schedule and budget hazards but often neglects the socially constructed nature of knowledge risks \cite{wallace2004software}.

In response, this paper introduces the KLRM Framework for software design and development. Recasting knowledge management focus areas as active risk controls, we integrate practices like code reviews, ADRs, and LLM-powered assistants into a systematic risk mitigation strategy \cite{skyrme1999knowledge, chau2003knowledge, fan2023large}. We argue that unmanaged knowledge risks are a hidden driver of failure in the iron triangle of scope, time, and cost \cite{atkinson1999project}. The KLRM framework addresses these by activating levers that compress rework cycles and accelerate onboarding. When properly governed, the LLM lever dramatically reduces rediscovery costs, freeing capacity for high-judgment architectural work \cite{fan2023large, peng2023impact}.

\section{Related Work}
The KLRM framework synthesizes concepts from intellectual capital theory, knowledge risk management (KRM), and emerging AI-augmented software engineering.

\subsection{Intellectual Capital in Software Engineering}
Intellectual Capital (IC) theory identifies that in knowledge-intensive domains, the most valuable assets walk out the door every evening \cite{stewart1997intellectual}. For software organizations, IC represents the expertise, processes, and relationships that produce and sustain the codebase. Following Bontis, we decompose IC into three dimensions: \cite{bontis1998intellectual}

\begin{itemize}
    \item \textbf{Human Capital (HC):} The programming skills, architectural expertise, and tacit debugging intuition of engineers. It is the most volatile asset, directly measured by the ``bus factor'' \cite{avelino2016novel}.
    \item \textbf{Structural Capital (SC):} Codified knowledge that persists independently of individuals: source code, ADRs, and automated pipelines. Failure to invest in SC leads to ``organizational amnesia'' \cite{rus2002knowledge, jansen2005software}.
    \item \textbf{Relational Capital (RC):} The value in external relationships with open-source communities, vendors, and API partners. Risks include dependency abandonment and vendor lock-in \cite{coelho2017why}.
\end{itemize}

The Knowledge-Based View (KBV) argues that a firm's primary purpose is the integration of specialized knowledge \cite{grant1996toward}. Despite this, Agile methodologies often provide limited guidance for managing the profound risks that arise when individuals---carrying critical tacit knowledge---depart the project \cite{durst2019mapping}.

\subsection{Knowledge Risks and Contextual Vulnerabilities}
KRM taxonomizes risks into human, operational, and technological domains \cite{durst2019mapping}. In SE, human risks involve attrition, knowledge hiding, and hoarding \cite{connelly2012knowledge}. Operational risks manifest as the reinvention of solutions due to poor retrospectives or onboarding. Technological risks include legacy decay and supply-chain vulnerabilities \cite{ilvonen2015towards, coelho2017why}.

Vulnerabilities vary by context. Agile teams generate high volumes of tacit knowledge that are rarely captured \cite{chau2003knowledge}. Distributed teams lose informal transfer mechanisms like overheard conversations. In open-source ecosystems, the ``paradox of openness'' creates relational risks where critical dependencies may be abandoned without warning \cite{coelho2017why}. Startups face hyper-concentrated risks where a single founder's departure can destroy product viability \cite{avelino2016novel}.

\subsection{The Rise of LLM-Augmented Software Development}
Large Language Models (LLMs) and AI coding assistants represent a paradigm shift in knowledge transfer \cite{fan2023large}. LLMs function as a cross-cutting lever: they amplify SC by auto-generating documentation and HC by enabling ``vibe coding'' where developers work above their expertise level \cite{peng2023impact, schaetzle2025research}.

However, this amplification introduces novel risks. \textit{Hallucination risk} contaminates structural capital with authoritative-looking falsehoods \cite{vaithilingam2022expectation}. \textit{Expertise atrophy} occurs as developers delegate cognitive work to AI, potentially hollowing out the human capital base required to sustain architectural integrity \cite{peng2023impact}. These dual dynamics necessitate explicit governance within any modern risk framework.

\subsection{Knowledge Levers as Risk Controls}
The concept of ``Knowledge Levers'' emerged from Skyrme and Dalkir as dynamic organizational focus areas that amplify value creation \cite{dalkir2011knowledge, skyrme2001capitalizing}. While previous literature treats these levers almost exclusively as tools for velocity and quality, the KLRM framework recasts them as instruments for systematic risk mitigation.

\section{The KLRM Framework for Software Engineering}
The goal of the KLRM Framework is to provide a structured, operational approach to mitigating knowledge-centric risks in software design and development. The foundational philosophy is that the systematic activation of software knowledge levers serves as the most effective defense against the taxonomy of knowledge risks identified in Section 2.

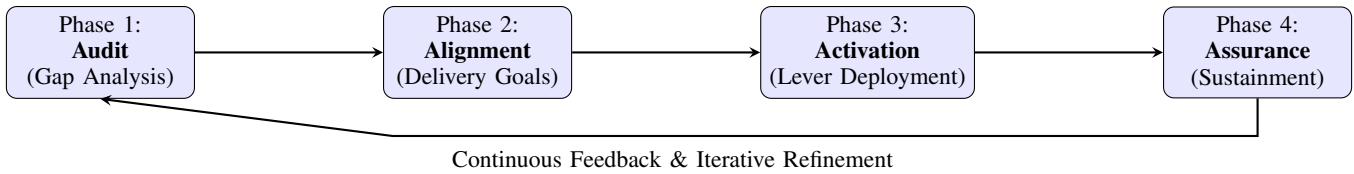
\begin{figure*}[t]
\centering
\begin{tikzpicture}[node distance=1.5cm, every node/.style={fill=white, font=\small}, align=center]
    \tikzstyle{phase} = [rectangle, rounded corners, minimum width=2.5cm, minimum height=1cm,text centered, draw=black, fill=blue!10]
    \tikzstyle{arrow} = [thick,->,>=stealth]
    
    \node (p1) [phase] {Phase 1: \\ \textbf{Audit} \\ (Gap Analysis)};
    \node (p2) [phase, right=of p1, xshift=1cm] {Phase 2: \\ \textbf{Alignment} \\ (Delivery Goals)};
    \node (p3) [phase, right=of p2, xshift=1cm] {Phase 3: \\ \textbf{Activation} \\ (Lever Deployment)};
    \node (p4) [phase, right=of p3, xshift=1cm] {Phase 4: \\ \textbf{Assurance} \\ (Sustainment)};

    \draw [arrow] (p1) -- (p2);
    \draw [arrow] (p2) -- (p3);
    \draw [arrow] (p3) -- (p4);
    \draw [arrow] (p4.south) -- ++(0,-0.5) -- ++(-11.5,0) -- (p1.south);
    
    \node at ($(p2.south)!0.5!(p3.south) + (0,-0.8)$) {Continuous Feedback \& Iterative Refinement};
\end{tikzpicture}
\caption{The KLRM Framework Architecture: Relationships between the four phases of knowledge risk mitigation.}
\label{fig:klrm_architecture}
\end{figure*}

The framework is structured across four sequential, iterative phases, as illustrated in Fig.~\ref{fig:klrm_architecture}.

\subsection{Phase 1: Software Knowledge Audit and Gap Analysis}
The KLRM framework begins with a comprehensive diagnostic of the software organization's knowledge landscape. Engineering managers must formally audit their knowledge assets to identify vulnerabilities before they manifest as project failures \cite{skyrme1999knowledge}.

This phase utilizes targeted diagnostic inquiries mapped to software engineering realities:
\begin{itemize}
    \item \textbf{Bus Factor Analysis:} For each critical system component, how many developers possess sufficient understanding to maintain, debug, and extend it? What is the team-level and organization-level bus factor \cite{avelino2016novel}? Tools such as Git-based code ownership analysis can quantify expertise concentration across modules.
    \item \textbf{Documentation Debt Assessment:} What percentage of architectural decisions lack recorded rationale? Are ADRs current, or have they drifted from the actual implementation? Is onboarding documentation sufficient for a new developer to become productive within a target time window \cite{jansen2005software, skyrme1999knowledge}?
    \item \textbf{Knowledge Silo Detection:} Are there specific teams, services, or infrastructure components where expertise is monopolized by a single developer or sub-team, effectively creating knowledge bottlenecks that block cross-functional collaboration \cite{skyrme1999knowledge}?
    \item \textbf{Retrospective Efficacy Audit:} Are Sprint retrospective action items being implemented and their impact measured, or have retrospectives degenerated into ritual activities producing no lasting organizational memory \cite{rus2002knowledge, skyrme1999knowledge}?
\end{itemize}

The outcome of Phase 1 is a knowledge risk heat map plotting knowledge criticality (how vital the expertise is to system reliability and development velocity) against the probability of loss or disruption. This translates the abstract concept of knowledge risk into a formalized risk register compatible with existing Agile risk management artifacts \cite{ilvonen2015towards}.

\subsection{Phase 2: Strategic Alignment with Delivery Objectives}
Knowledge management initiatives in software organizations frequently fail when decoupled from engineering delivery goals, appearing to developers as administrative overhead that competes with ``real work.'' Phase 2 aligns identified knowledge risks with the organization's delivery objectives and engineering strategy.

This requires establishing direct, communicable links between specific knowledge levers and targeted engineering outcomes \cite{skyrme1999knowledge}. For example, if a platform team's primary objective is to reduce mean time to recovery (MTTR) for production incidents, and the audit reveals that incident response knowledge is concentrated in two senior SREs(Site Reliability Engineers) with no documented runbooks, the strategy must explicitly prioritize the ``Organizational Memory'' and ``Knowledge in Processes'' levers. Quick wins---such as documenting the top-10 most frequent incident resolution procedures---demonstrate the KLRM approach's value, justify further investment, and begin the cultural shift toward knowledge-sharing practices \cite{dalkir2011knowledge, skyrme1999knowledge}.

\subsection{Phase 3: Activating Knowledge Levers as Risk Controls}
The operational core of the KLRM framework deploys eight knowledge levers as engineered countermeasures against software knowledge risks \cite{durst2019mapping}.

\begin{itemize}
    \item \textbf{1. Developer Expertise (People Lever):} Mitigates bus-factor and knowledge hiding risks. Activation includes pair programming rotations, structured mentorship, and tech talks to transform tacit individual expertise into shared team intelligence \cite{connelly2012knowledge, skyrme2001capitalizing, chau2003knowledge}.
    \item \textbf{2. Organizational Memory (Structural Capital):} Targets design rationale loss. Activation involves mandating ADRs, searchable post-mortem databases, and ``documentation as code'' to preserve design intent independent of personnel \cite{skyrme2001capitalizing, jansen2005software}.
    \item \textbf{3. Knowledge in Processes:} Addresses quality drift. Activation embeds knowledge into CI/CD pipelines (linting, automated tests) and structures retrospectives as formal knowledge extraction events \cite{skyrme2001capitalizing, chau2003knowledge}.
    \item \textbf{4. Ecosystem Relationships (Relational Capital):} Targets open-source dependency risk, vendor lock-in, outsourcing knowledge leakage, and API partner instability. Activation involves establishing formal dependency health monitoring (tracking maintainer activity, vulnerability disclosure patterns, license changes), creating vendor evaluation frameworks, and implementing structured knowledge transfer protocols with outsourcing partners \cite{coelho2017why}.
    \item \textbf{5. User and Product Knowledge:} Targets misalignment between engineering effort and user needs, feature irrelevance, and wasted development cycles. Activation involves integrating product telemetry, user research findings, and customer support patterns directly into sprint planning, creating feedback loops that ensure engineering decisions are grounded in empirical user behavior rather than assumptions \cite{skyrme2001capitalizing}.
    \item \textbf{6. Knowledge in Products:} Targets operational opacity, debugging difficulty, and the loss of runtime behavioral understanding. Activation involves embedding observability instrumentation (structured logging, distributed tracing, health check endpoints), self-documenting API conventions, and comprehensive error messages that encode diagnostic knowledge for future operators \cite{skyrme2001capitalizing}.
    \item \textbf{7. Technology Radar (Environmental Insights):} Targets technology obsolescence, ecosystem disruption, and strategic misalignment with industry trends. Activation involves maintaining an internal technology radar, running periodic architecture review boards, and establishing RFC processes for evaluating new frameworks, languages, or infrastructure components \cite{skyrme2001capitalizing}.
    \item \textbf{8. AI-Augmented Development (LLM Lever):} Targets documentation scarcity, onboarding delays, legacy code incomprehensibility, and knowledge bottlenecks during incident response. Activation involves deploying LLM-based coding assistants for AI pair programming, using LLMs to auto-generate and maintain documentation from codebases, leveraging agentic systems for automated code review and refactoring, and enabling natural-language querying of organizational knowledge bases. Critically, activation \textit{also} requires governance countermeasures: mandatory human review of AI-generated artifacts, automated validation of LLM outputs against actual codebase behavior, and deliberate ``deep work'' sessions where developers engage with systems without AI assistance to maintain expertise depth \cite{schaetzle2025research, fan2023large, peng2023impact, vaithilingam2022expectation}.
\end{itemize}

\subsubsection*{Granular Mechanisms of Lever Activation}

\textbf{Mitigating Bus-Factor Risk via the Developer Expertise Lever:}
Bus-factor risk is the most acute and most neglected knowledge risk in software engineering. The KLRM framework addresses it by targeting the incentive structures and cultural norms that enable expertise concentration. Knowledge hiding in software teams is often a rational response to cultures that reward individual heroics---the developer who is the ``only one who can fix production at 3 AM'' receives disproportionate recognition, inadvertently incentivizing expertise hoarding \cite{connelly2012knowledge}. The Developer Expertise lever is activated by implementing systematic pair programming rotations (ensuring every critical system component has at least three developers with operational understanding), integrating knowledge-sharing metrics into performance evaluations (rewarding developers who write ADRs, mentor juniors, or conduct knowledge transfer sessions), and fostering a culture of psychological safety where asking questions and admitting knowledge gaps is normalized \cite{skyrme2001capitalizing, chau2003knowledge}.

\textbf{Mitigating Open-Source Dependency Risk via the Ecosystem Relationships Lever:}
Modern software projects typically depend on hundreds to thousands of open-source packages, creating a form of relational capital risk unique to software engineering \cite{coelho2017why}. The KLRM framework deploys the Ecosystem Relationships lever to systematically monitor dependency health, evaluate maintainer responsiveness and community vibrancy before adopting new dependencies, maintain internal forks or abstraction layers for critical dependencies, and establish contributing relationships with strategically important open-source projects to ensure organizational influence over their direction \cite{coelho2017why}.

\textbf{Mitigating Operational Amnesia via Organizational Memory and Process Levers:}
In software organizations, the failure to institutionalize knowledge from production incidents, architectural decisions, and sprint retrospectives leads to chronic ``reinvention of the wheel''---teams debugging issues that were solved months ago by another team, or making architectural decisions that contradict lessons learned from previous project failures \cite{rus2002knowledge}. The KLRM framework mandates the extraction of tacit knowledge generated during incidents and projects into explicit organizational memory (post-mortem databases, ADRs, runbooks) \cite{jansen2005software}. This structural capital is then operationalized through the Process lever: past incident patterns become automated monitoring alerts, architectural anti-patterns become PR review checklist items, and retrospective insights become engineering process improvements tracked in a dedicated backlog.

\textbf{The LLM Lever: Amplifier and Risk Source:}
The AI-Augmented Development lever occupies a unique position within the KLRM framework: it is simultaneously the most powerful knowledge amplifier available to modern software teams and a source of novel, poorly understood risks that require deliberate governance.

\textit{Knowledge amplification benefits.} LLM-based coding assistants function as a ``knowledge multiplier'' that accelerates every other lever in the framework. When paired with the Organizational Memory lever, LLMs can auto-generate ADRs from code review discussions, produce runbooks from incident chat logs, and maintain living documentation that stays synchronized with the codebase---dramatically lowering the marginal cost of structural capital creation \cite{schaetzle2025research, fan2023large}. When paired with the Developer Expertise lever, AI pair programming enables junior developers to work productively on unfamiliar systems by providing contextual explanations, suggesting idiomatic patterns, and surfacing relevant documentation on demand---effectively compressing the onboarding timeline \cite{peng2023impact}. Agentic coding systems go further, autonomously executing multi-step development tasks (code generation, testing, refactoring, deployment) that previously required deep specialized expertise \cite{fan2023large}. For under-documented legacy systems---a chronic pain point in software organizations---LLMs can reconstruct design rationale by analyzing commit histories, code structure, and test patterns, partially reversing the architectural knowledge vaporization described in Section 1 \cite{jansen2005software, fan2023large}.

\textit{Knowledge risks introduced.} However, the very capabilities that make LLMs powerful also create three categories of novel risk:

\begin{enumerate}
    \item \textbf{Hallucination risk (structural capital contamination):} LLMs generate plausible but incorrect code, documentation, and API contracts with high confidence. When these artifacts enter the organizational memory without rigorous human validation, they contaminate structural capital with authoritative-looking falsehoods---a category of knowledge risk with no historical precedent in software engineering \cite{schaetzle2025research, vaithilingam2022expectation}. Unlike human errors, which tend to be obviously wrong or internally inconsistent, LLM hallucinations are designed by architecture to be maximally plausible, making them significantly harder to detect through casual review.
    \item \textbf{Expertise atrophy (human capital erosion):} As developers increasingly delegate cognitive work to AI assistants, their deep understanding of system architecture, debugging strategies, and design trade-offs may degrade over time \cite{vaithilingam2022expectation}. The phenomenon of ``vibe coding''---generating functional code through iterative prompting without fully comprehending the underlying logic---exemplifies this risk. While it accelerates short-term delivery, it may hollow out the human capital base ($H$) that the mathematical model identifies as the single most critical driver of knowledge capital ($K$). If $H$ declines due to expertise atrophy, the $\beta \cdot H$ codification coupling means that structural capital creation also slows, creating a compounding degradation loop \cite{peng2023impact}.
    \item \textbf{Overreliance and single-point-of-failure risk:} Organizations that embed LLM assistants deeply into their development workflows create a new form of dependency risk. Service outages, pricing changes, or policy shifts by LLM providers can suddenly disrupt development velocity---analogous to the open-source dependency risk described in Section 2.3, but affecting human productivity rather than software components \cite{vaithilingam2022expectation}. Teams that have adapted their processes around AI assistance may find their effective expertise reduced when that assistance is unavailable.
\end{enumerate}

\textit{Governance mechanisms.} The KLRM framework addresses these risks through a balanced activation strategy: organizations should aggressively leverage LLMs for knowledge amplification while implementing explicit governance countermeasures. These include: (1) mandatory human review gates for all AI-generated documentation and code entering the organizational memory; (2) automated validation pipelines that test LLM-generated code against actual system behavior (e.g., running generated tests, verifying API contracts against live endpoints); (3) ``deep understanding'' time---dedicated engineering hours where developers work without AI assistance on core system components to maintain and deepen their architectural understanding; (4) LLM output provenance tracking that labels AI-generated artifacts so that future consumers can apply appropriate skepticism; and (5) multi-provider strategies that prevent single-vendor dependency on any one LLM platform \cite{schaetzle2025research, vaithilingam2022expectation}.

\subsection{Phase 4: Assurance, Sustainment, and Capability Building}
The final phase ensures that knowledge risk management becomes a persistent capability embedded within the engineering culture. This requires defining specific roles responsible for knowledge stewardship---whether a dedicated Knowledge Engineering lead, rotating ``documentation champions,'' or engineering managers with explicit KRM objectives in their performance criteria \cite{chau2003knowledge}.

Organizations must continuously measure lever impact using engineering metrics aligned with their delivery objectives: DORA metrics (deployment frequency, lead time, change failure rate, MTTR), onboarding time-to-productivity, bus factor trends, documentation coverage ratios, and retrospective action item completion rates \cite{forsgren2018accelerate}. By treating KRM as a core component of engineering excellence rather than an administrative side project, organizations can dynamically adjust lever activation in response to emerging risks---such as a sudden spike in developer turnover or the deprecation of a critical dependency \cite{skyrme1999knowledge}.

\section{Mathematical Model and Simulation}
This section provides a formal quantitative test of the KLRM framework. We use a mathematical model to simulate how knowledge grows and disappears in a software project, comparing a project with no levers (Baseline) to one with full lever activation.

The model relies on three fundamental assumptions to mirror real-world software engineering. First, we treat knowledge as a \textbf{stochastic (random) process} because project outcomes are never deterministic; they are subject to constant uncertainty. Second, we use \textbf{Poisson processes} to model ``shocks'' like developer attrition or system outages. This is because such events are discrete, sudden, and occur independently over time---we know their average frequency, but never their exact timing. Finally, we model learning and documentation decay as \textbf{continuous flows}, reflecting the reality that while a developer quits in an instant, the gradual accumulation of expertise and the slow ``rot'' of outdated documentation happen every day.

\subsection{Model Formulation}
We represent a project's ``Knowledge Capital''---the operationalized equivalent of its Intellectual Capital---as a composite score ($K$) derived from three parts:
\begin{itemize}
    \item \textbf{Human Capital ($H$):} Developer expertise and intuition.
    \item \textbf{Structural Capital ($S$):} Documentation, code, and ADRs.
    \item \textbf{Relational Capital ($R$):} External community and vendor trust.
\end{itemize}

The total score at any time $t$ is calculated as:
\begin{equation}
    \label{eq:composite_index}
    K(t) = 0.40 \cdot H(t) + 0.35 \cdot S(t) + 0.25 \cdot R(t)
\end{equation}
The weights in Eq. \eqref{eq:composite_index} are illustrative for simulation purposes; alternative values can be used to reflect different project priorities and contexts.

\subsubsection*{The Dynamics---Growth, Decay, and Shocks}
The model tracks how $H, S,$ and $R$ change over time. In simple terms, the change in knowledge follows a three-part logic:
\begin{center}
    \textit{[Change] = [Growth] $-$ [Natural Decay] $-$ [Sudden Shocks]}
\end{center}

The formal stochastic differential equations governing these dynamics are:
\begin{align}
    \frac{dH}{dt} &= \alpha_H - \delta_H \cdot H(t) - J_H \cdot \frac{dN_H(t)}{dt} \label{eq:dH}\\[6pt]
    \frac{dS}{dt} &= \beta \cdot H(t) - \gamma_S \cdot S(t) - J_S \cdot \frac{dN_S(t)}{dt} \label{eq:dS}\\[6pt]
    \frac{dR}{dt} &= \alpha_R - \delta_R \cdot R(t) - J_R \cdot \frac{dN_R(t)}{dt} \label{eq:dR}
\end{align}
where $\alpha_H, \alpha_R$ are growth rates, $\delta_H, \delta_R, \gamma_S$ are decay rates, $\beta$ is the human-to-structural codification coupling, $J_i$ are shock magnitudes, and $N_i(t)$ are Poisson processes governing discrete shock arrivals.

Key behavioral rules in the model include:
\begin{itemize}
    \item \textbf{Natural Decay:} Knowledge "leaks" over time. Documentation becomes outdated ($\gamma_S$), and people forget details ($\delta_H$).
    \item \textbf{The Human-to-Structural Link:} In Eq.~\eqref{eq:dS}, structural capital ($S$) can only grow if human expertise ($H$) is present. This captures the reality that you cannot write high-quality documentation for a system you do not understand \cite{jansen2005software}.
    \item \textbf{Sudden Shocks:} These represent catastrophic events, such as a lead architect quitting unexpectedly or a critical third-party API being shut down. These events cause an immediate drop in the knowledge score.
\end{itemize}

\subsubsection*{How Knowledge Levers Work}
Activating a Knowledge Lever ($\lambda$) helps the project in two specific ways:
\begin{enumerate}
    \item \textbf{Boosting Growth:} Levers act like a "volume knob" for knowledge creation. For example, the Developer Expertise lever ($\lambda_P$) increases the rate of skill development through mentorship and pair programming.
    \item \textbf{Cushioning Losses:} Levers act as "shock absorbers." If a developer leaves, a project with an active Organizational Memory lever ($\lambda_M$) loses significantly less value because their knowledge was already codified into ADRs and runbooks.
\end{enumerate}

\subsection{Simulation Design}
To test the model, we use Monte Carlo simulations. This involves playing out 5,000 different possible 10-year futures for a project. Each future starts with the same conditions but experiences random "shocks" at different times. By looking at the average of all 5,000 scenarios, we can see the true impact of the levers.

We measure the results using:
\begin{itemize}
    \item \textbf{Expected Capital:} The average knowledge score after 10 years.
    \item \textbf{Sharpe Ratio:} A measure of "Value-for-Risk" \cite{sharpe1966mutual}. A higher Sharpe ratio means the project is not just more knowledgeable, but also more stable and predictable.
    \item \textbf{Crisis Probability:} The chance that the knowledge score drops so low ($<40$) that the project is likely to fail.
\end{itemize}

Six scenarios isolate individual and combined lever effects:
\begin{enumerate}
    \item \textbf{Baseline} (all $\lambda = 0$): No lever activation
    \item \textbf{Full KLRM} ($\lambda_P = 0.6,\; \lambda_M = 0.6,\; \lambda_{Pr} = 0.5,\; \lambda_R = 0.5$): All levers active
    \item \textbf{Developer Expertise Only} ($\lambda_P = 0.6$)
    \item \textbf{Organizational Memory Only} ($\lambda_M = 0.6$)
    \item \textbf{Process Only} ($\lambda_{Pr} = 0.5$)
    \item \textbf{Ecosystem Relationships Only} ($\lambda_R = 0.5$)
\end{enumerate}

Metrics: expected terminal knowledge capital $\mathbb{E}[K(T)]$, coefficient of variation $\text{CV} = \sigma / \mathbb{E}$, knowledge Sharpe ratio $\mathbb{E} / \sigma$ \cite{sharpe1966mutual}, and crisis probability $\mathbb{P}(K < K^*)$ where $K^* = 40$.

\subsection{Results}
Table~\ref{tab:results} summarizes the terminal-state statistics. The divergent trajectories of project knowledge capital under different governance regimes are illustrated in Figure~\ref{fig:sample_paths}.

\begin{table}[ht]
\centering
\caption{Monte Carlo simulation results ($N = 5{,}000$ paths, $T = 10$ years)}
\label{tab:results}
\footnotesize
\begin{tabular}{lcccc}
\toprule
\textbf{Scenario} & $\mathbb{E}[K(T)]$ & $\text{CV}$ (\%) & \textbf{Sharpe} & $\mathbb{P}(\text{crisis})$ (\%) \\
\midrule
Baseline                 & 53.35 & 10.3 &  9.73 & 0.64 \\
Dev. Expertise Only      & 68.19 &  8.6 & 11.65 & 0.00 \\
Org. Memory Only         & 59.32 & 10.1 &  9.91 & 0.02 \\
Process Only             & 58.30 & 10.4 &  9.64 & 0.10 \\
Ecosystem Rel.           & 58.15 &  9.4 & 10.60 & 0.06 \\
\textbf{Full KLRM}       & \textbf{87.39} & \textbf{7.7} & \textbf{12.99} & \textbf{0.00} \\
\bottomrule
\end{tabular}
\end{table}

\subsubsection*{Key Findings}
\begin{figure*}[ht]
    \centering
    \includegraphics[width=\textwidth]{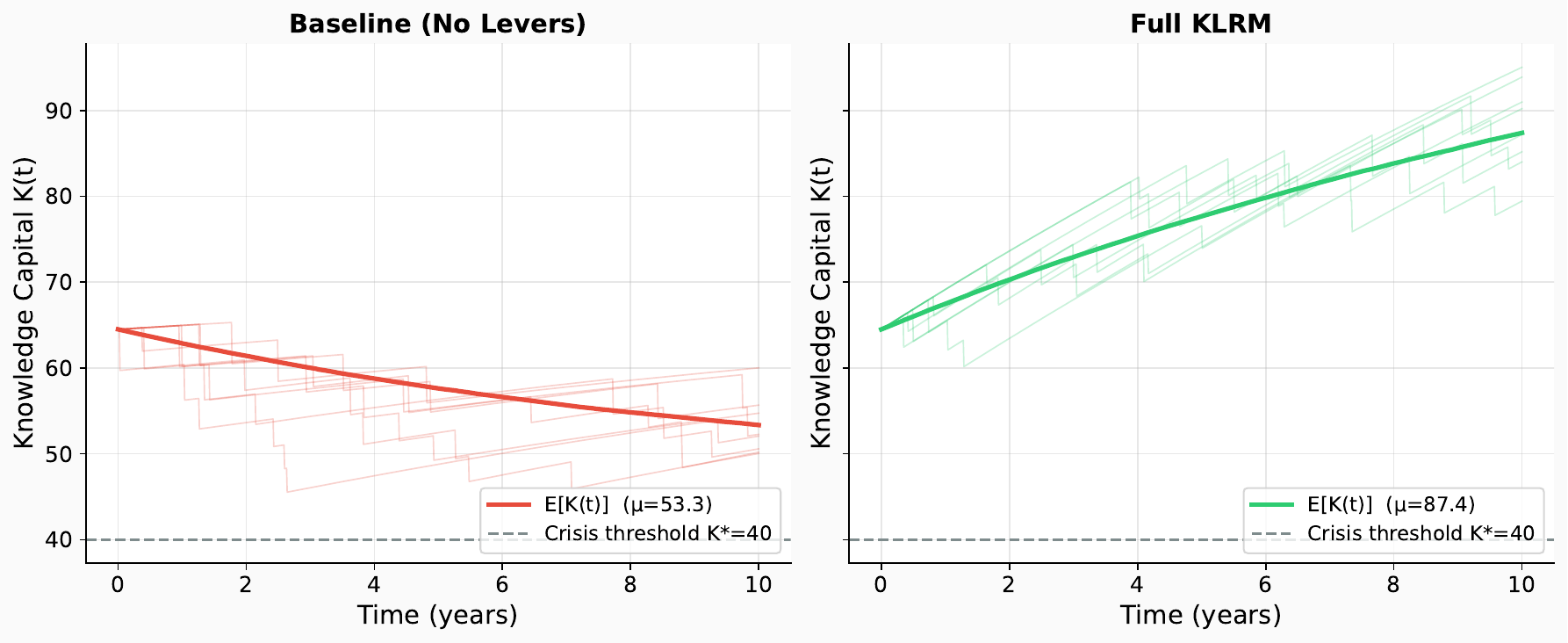}
    \caption{Sample paths of software project knowledge capital $K(t)$. Left: Baseline (no levers). Right: Full KLRM. The ensemble mean $\mathbb{E}[K(t)]$ (bold) shows the divergent trajectories; the dashed line marks the crisis threshold $K^* = 40$. Lever activation stabilizes the stochastic decay and prevents catastrophic knowledge loss.}
    \label{fig:sample_paths}
\end{figure*}

\textbf{Finding 1: Lever activation significantly increases expected knowledge capital.} Full KLRM produces $\mathbb{E}[K(T)] = 87.39$, a $+63.8\%$ improvement over Baseline. Figure~\ref{fig:sample_paths} (right) shows how lever activation shifts the ensemble mean upward while tightening the variance around the mean compared to the Baseline (left). This effect is superadditive: the Developer Expertise lever increases $H$, which amplifies the Organizational Memory lever's $H \rightarrow S$ codification pathway.

\begin{figure}[ht]
    \centering
    \includegraphics[width=\columnwidth]{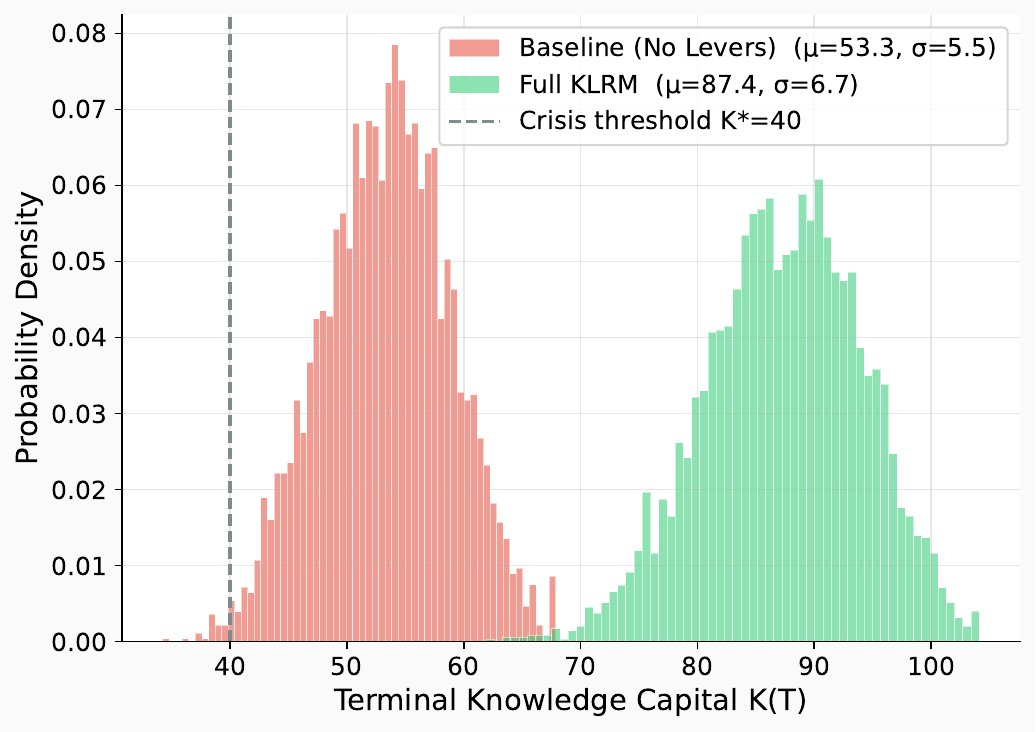}
    \caption{Terminal distribution of $K(T)$ at $T = 10$ years. The Full KLRM distribution (green) is shifted rightward with lower relative spread, confirming reduced coefficient of variation and increased reliability of project capability.}
    \label{fig:terminal_dist}
\end{figure}

\textbf{Finding 2: Risk-adjusted volatility is substantially reduced.} The coefficient of variation decreases by 25.2\% under Full KLRM. This rightward shift and narrowing of the knowledge distribution is clearly visible in the terminal distribution plot (Figure~\ref{fig:terminal_dist}), where the Full KLRM distribution (green) exhibits a higher mean and lower spread than the Baseline.

\begin{figure}[ht]
    \centering
    \includegraphics[width=\columnwidth]{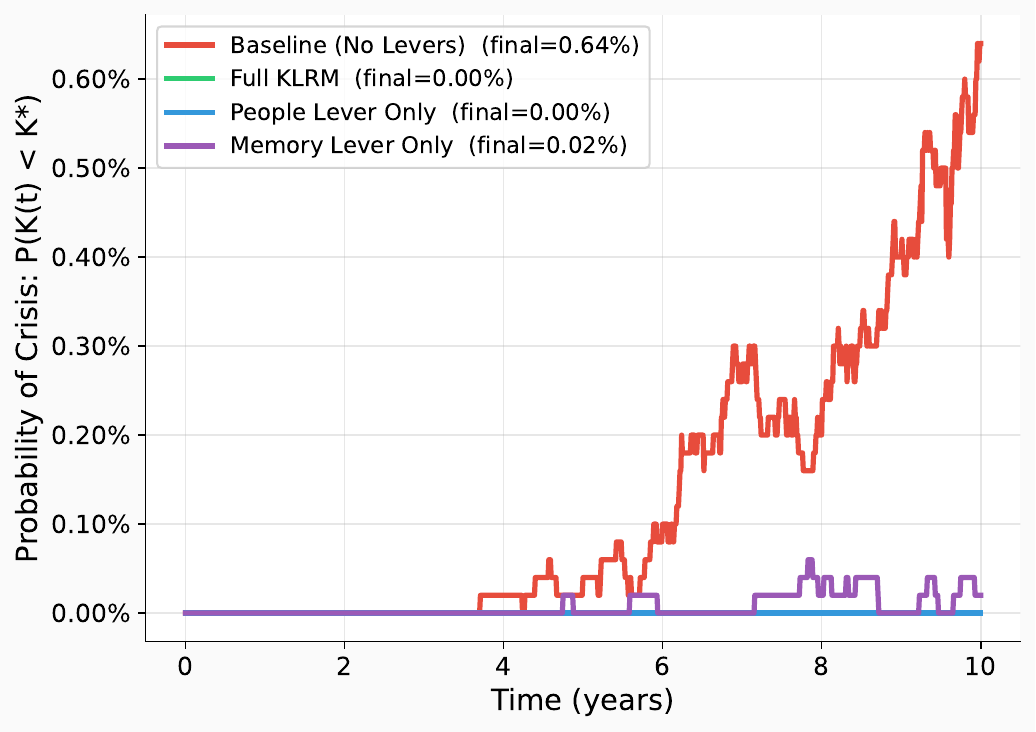}
    \caption{Knowledge crisis probability $\mathbb{P}(K < K^*)$ over time. Lever-activated scenarios maintain near-zero crisis probability throughout the simulation horizon, effectively eliminating the risk of project failure due to knowledge scarcity.}
    \label{fig:crisis_prob}
\end{figure}

\textbf{Finding 3: Knowledge crisis probability is virtually eliminated.} The Baseline crisis probability of 0.64\% is effectively mitigated under lever activation. As shown in Figure~\ref{fig:crisis_prob}, the probability of $K(t)$ falling below the critical threshold $K^*$ remains near zero throughout the 10-year horizon for all lever-activated scenarios, whereas the Baseline shows persistent vulnerability.

\begin{figure*}[ht]
    \centering
    \includegraphics[width=\textwidth]{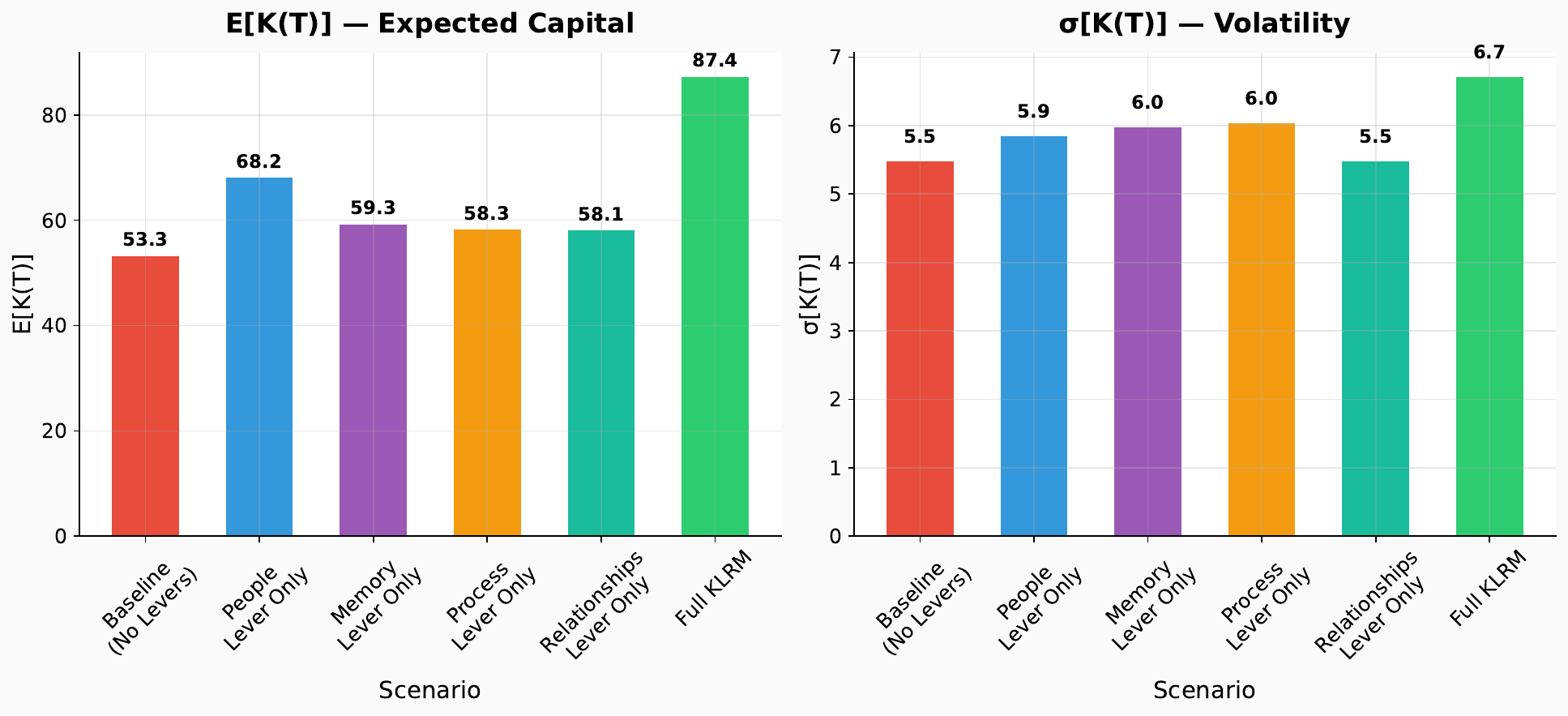}
    \caption{Lever decomposition: $\mathbb{E}[K(T)]$ (left) and $\sigma[K(T)]$ (right) by scenario, demonstrating the superadditive effect of combined lever activation. Human-centric levers provide the foundational stability required for structural levers to be effective.}
    \label{fig:lever_decomp}
\end{figure*}

\textbf{Finding 4: Developer Expertise remains the primary driver of capability.} Among individual levers, Developer Expertise produces the largest improvement ($+27.8\%$). The relative contribution of each lever to both terminal capital level and stability is decomposed in Figure~\ref{fig:lever_decomp}, which highlights the dominance of people-centric interventions in sustaining software R\&D capability.

\subsection{Discussion}
The simulation results provide formal quantitative support for the KLRM framework's cnoceptual model. Three contributions emerge.

First, knowledge levers function as \textit{structural stabilizers} for software projects---they increase expected knowledge capital while reducing relative volatility and crisis probability. This dual effect is analogous to portfolio diversification in financial risk management \cite{sharpe1966mutual} and provides a rigorous justification for treating practices like pair programming, ADRs, and code review not merely as ``good engineering hygiene'' but as formal risk controls that belong in the project risk register.

Second, the superadditive People-Memory coupling (mediated by $\beta \cdot H$) validates the framework's emphasis on activating levers in concert. Organizations that invest in documentation infrastructure (Organizational Memory lever, $\mathbb{E}[K(T)] = 59.32$) without simultaneously developing the developer expertise to populate it ($68.19$ for Developer Expertise alone) receive diminished returns. This formalizes the common SE observation that ``tools don't write documentation; people do.''

Third, the quantitative dominance of the Developer Expertise lever provides formal support for the prioritization of people-centric interventions---pair programming, mentorship, and knowledge-sharing cultures---as the highest-leverage risk mitigation strategy for resource-constrained software teams.

Fourth, and most consequentially for software project management practice, the simulation results illuminate the mechanism through which knowledge risks silently degrade iron triangle performance \cite{atkinson1999project}. In the Baseline scenario, knowledge capital declines and exhibits high volatility; translated to project management terms, this means that scope estimation becomes increasingly unreliable (because fewer people understand system complexity), schedules slip (because developers spend time rediscovering undocumented design rationale rather than building features), and costs inflate (because knowledge-poor teams require more rework cycles, longer debugging sessions, and slower onboarding of new staff). The Full KLRM scenario's 63.8\% increase in $\mathbb{E}[K(T)]$ and 25.2\% reduction in CV represent, in iron triangle terms, a workforce that estimates more accurately (scope), executes more predictably (time), and wastes less effort on rediscovery and rework (cost).

The LLM lever plays a particularly pivotal role in this iron triangle alignment. By compressing onboarding timelines (junior developers become productive faster with AI assistance), reducing documentation debt (auto-generated ADRs and runbooks lower the marginal cost of structural capital), and accelerating routine implementation (agentic systems handle boilerplate, freeing developers for high-judgment architectural work), the LLM lever directly attacks the knowledge-driven inefficiencies that cause software projects to exceed their budgets and miss their deadlines \cite{fan2023large, peng2023impact}. However, the governance requirements identified in Section 3---human review gates, expertise atrophy countermeasures, and multi-provider strategies---are essential: an improperly governed LLM lever risks creating a false sense of velocity (delivering code that ``works'' but embodies hallucinated assumptions, generating rework downstream) and deepening the expertise atrophy that makes accurate scope estimation impossible \cite{vaithilingam2022expectation}. The KLRM framework thus provides a structured approach for realizing the iron triangle benefits of AI-assisted development while managing the novel risks it introduces.

This model is a stylized representation; parameters are calibrated to plausible ranges rather than derived from empirical data. Future calibration using repository mining and DORA metrics data would strengthen prescriptive utility (Section 6).

\section{Future Work}
The KLRM framework, including the AI-Augmented Development lever, establishes a theoretical and operational baseline for managing knowledge risks in software projects. Future research should pursue three trajectories.

\subsection{Empirical Measurement of LLM Knowledge Effects}
While this paper integrates LLMs as a core knowledge lever and identifies their dual amplification-risk dynamics, empirical validation remains urgently needed. Future work should design controlled experiments measuring: (1) the rate at which LLM-generated documentation introduces factual errors into organizational memory (hallucination contamination rate); (2) the longitudinal impact of AI-assisted development on developer expertise depth (expertise atrophy measurement using code comprehension assessments over 6--12 month periods); (3) the net effect of LLM adoption on bus factor---whether AI pair programming genuinely distributes knowledge or merely enables developers to operate in systems they do not deeply understand; and (4) the productivity cliff experienced when LLM access is interrupted, quantifying organizational dependency risk \cite{schaetzle2025research, fan2023large, peng2023impact, vaithilingam2022expectation}. Such studies would enable empirical calibration of the knowledge amplification vs. knowledge risk trade-off that defines the LLM lever's governance requirements.

\subsection{Empirical Calibration Using Software Engineering Metrics}
The current model uses stylized parameters. Future work should calibrate the model empirically using real-world software engineering data: DORA metrics (deployment frequency, lead time, change failure rate, MTTR) as proxy measures for $K(t)$; Git repository mining to quantify bus factor dynamics ($H$ concentration); documentation coverage tools to measure $S$ evolution; and dependency health metrics to track $R$ volatility \cite{forsgren2018accelerate, kruchten2012technical}. Longitudinal studies correlating KLRM lever activation with measurable improvements in software delivery performance would provide the empirical grounding needed to elevate KRM from a theoretical framework to a practitioner-validated methodology.

\subsection{Context-Specific Validation Across Software Development Paradigms}
Knowledge risk profiles differ dramatically across software development contexts. Enterprise software teams managing long-lived monolithic systems face knowledge risks dominated by legacy comprehension and architectural debt, while fast-moving startup teams face risks concentrated in bus-factor exposure and documentation scarcity \cite{avelino2016novel}. Open-source projects present entirely different challenges centered on maintainer burnout, contributor onboarding, and community governance \cite{coelho2017why}.

Future research should conduct comparative case studies examining how lever weighting must be adjusted for enterprise vs. startup vs. open-source contexts, and how Agile-specific practices (retrospectives, stand-ups, sprint reviews) can be systematically enhanced to function as knowledge risk controls rather than purely ceremonial process artifacts \cite{rus2002knowledge, chau2003knowledge}.

\bibliographystyle{IEEEtran}
\bibliography{references}

\end{document}